# A NON LINEAR APPROACH TOWARDS AUTOMATED EMOTION ANALYSIS IN HINDUSTANI MUSIC

Shankha Sanyal*[1,2], Archi Banerjee[1,2], Tarit Guhathakurata[1],
Ranjan Sengupta[1] and Dipak Ghosh[1]

[1]Sir C.V. Raman Centre for Physics and Music, [2]Department of Physics
Jadavpur University, Kolkata: 700032
*ssanyal.sanyal2@gmail.com

**ABSTRACT:**
*In North Indian Classical Music, raga forms the basic structure over which individual improvisations is performed by an artist based on his/her creativity. The Alap is the opening section of a typical Hindustani Music (HM) performance, where the raga is introduced and the paths of its development are revealed using all the notes used in that particular raga and allowed transitions between them with proper distribution over time. In India, corresponding to each raga, several emotional flavors are listed, namely erotic love, pathetic, devotional, comic, horrific, repugnant, heroic, fantastic, furious, peaceful. The detection of emotional cues from Hindustani Classical music is a demanding task due to the inherent ambiguity present in the different ragas, which makes it difficult to identify any particular emotion from a certain raga. In this study we took the help of a high resolution mathematical microscope (MFDFA or Multifractal Detrended Fluctuation Analysis) to procure information about the inherent complexities and time series fluctuations that constitute an acoustic signal. With the help of this technique, 3 min alap portion of six conventional ragas of Hindustani classical music namely, Darbari Kanada, Yaman, Mian ki Malhar, Durga, Jay Jayanti and Hamswadhani played in three different musical instruments were analyzed. The results are discussed in detail.*

# INTRODUCTION:

Musical instruments are often thought of as linear harmonic systems, and a first-order description of their operation can indeed be given on this basis. The term 'linear' implies that an increase in the input simply increases the output proportionally, and the effects of different inputs are only additive in nature. The term 'harmonic' implies that the sound can be described in terms of components with frequencies that are integral multiples of some fundamental frequency, which is essentially an approximation and the reality is quite different. Most of the musical instruments have resonators that are only approximately harmonic in nature, and their operation and harmonic sound spectrum both rely upon the extreme nonlinearity of their driving mechanisms. Such instruments might be described as 'essentially nonlinear' [2].

The three instruments chosen for our analysis are *sitar, sarod and flute*. All of them have been phenomenal for the growth and spread of Hindustani classical music over the years. The first two are plucked string instruments having a non-linear bridge structure, which is what gives them a very distinct characteristic buzzing timbre. It has been shown in earlier studies that the mode frequencies of a real string are not exactly harmonic, but relatively stretched because of stiffness [1], and that the mode frequencies of even simple cylindrical pipes are very appreciably inharmonic because of variation of the end correction with frequency; hence a non linear treatment of the musical signals generated from these instruments become invincible. Non-linear fractal analysis/physical modeling of North Indian musical instruments were done in a few earlier works [2-4]; but using them to quantify and categorize emotional appraisal has never been done before. That music has its effect in triggering a multitude of reactions on the human brain is no secret. However, there has been little scientific investigation in the Indian context [5,6] on whether different moods are indeed elicited by different *ragas* and how they depend on the underlying structure of the *raga*.

We chose 3 min *alap* portion of six conventional *ragas* of Hindustani classical music namely, *"Darbari Kanada", "Yaman", "Mian ki Malhar", "Durga", "Jay Jayanti" and "Hamswadhani"* played in three different musical instruments. The first three *ragas* correspond to the negative dimension of the Russel's emotional sphere, while the last three belong to the positive dimension (conventionally).

---

\* Corresponding Author

The music signals were analyzed with the help of latest non linear analysis technique called Multifractal Detrended Fluctuation Analysis (MFDFA) [7] which determines the complexity parameters associated with each *raga* clips. With the help of this technique, we have computed the multifractal spectral width (or the complexity) associated with each *raga* clip. The complexity values give clear indication in the direction of categorization of emotions attributed to Hindustani classical music as well as timbre specification of a particular instrument. The inherent ambiguities present in each *raga* of Hindustani classical music is also beautifully reflected in the results. The complexity value corresponding to different parts of a particular raga becomes almost similar to the values corresponding to parts of a different *raga*. This implies acoustic similarities in these parts and hence the emotional attributes of these parts are bound to be similar. In this way, we have tried to develop automated algorithm with which we can classify and quantify emotional arousal corresponding to different *ragas* of Hindustani music.

## EXPERIMENTAL DETAILS:

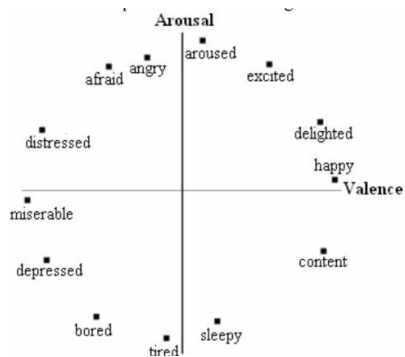

Fig. 1: Russel's arousal valence 2D-model of emotion

Six different *ragas* of Hindustani Classical music played in traditional flute, *sitar* and *sarod* were taken for our analysis. The *ragas* were chosen by an experienced musician such that they belong to the positive and negative valence of the 2D emotional sphere illustrated in **Fig. 1**. We chose 3 min *alap* portion of six conventional *ragas* of Hindustani classical music namely, "*Darbari Kanada*", "*Yaman*", "*Mian ki Malhar*", "*Durga*", "*Jay Jayanti*" and "*Hamswadhani*" played in three different musical instruments. The signals are digitized at the rate of 22050 samples/sec 16 bit format. The *alaap* part characteristic features of the entire *raga* is present in this part and that it uses all the notes used in that particular raga and allowed transitions between them with proper distribution over time. Each three minutes signal is divided into four equal segments of 45 seconds each. We measured the multifractal spectral width (or the complexity) corresponding to each of the 45 second fragments of the Hindustani *raga*.

## METHOD OF ANALYSIS

**Method of multifractal analysis of sound signals**

The time series data obtained from the sound signals are analyzed using MATLAB [8] and for each step an equivalent mathematical representation is given which is taken from the prescription of Kantelhardt et al [7].

The complete procedure is divided into the following steps:

*Step 1:* Converting the noise like structure of the signal into a random walk like signal. It can be represented as:

$$Y(i) = \sum (x_k - \bar{x}) \quad (1)$$

Where $\bar{x}$ is the mean value of the signal.

*Step 2:* The local RMS variation for any sample size *s* is the function *F(s,v)*. This function can be written as follows:

$$F^2(s,v) = \frac{1}{s} \sum_{i=1}^{s} \{Y[(v-1)s+i] - y_v(i)\}^2$$

*Step 4:* The q-order overall RMS variation for various scale sizes can be obtained by the use of following equation

$$F_q(s) = \left\{ \frac{1}{Ns} \sum_{v=1}^{Ns} [F^2(s,v)]^{\frac{q}{2}} \right\}^{\left(\frac{1}{q}\right)} \quad (2)$$

*Step 5:* The scaling behaviour of the fluctuation function is obtained by drawing the log-log plot of $F_q(s)$ vs. s for each value of q.

$$F_q(s) \sim s^{h(q)} \quad (3)$$

The h(q) is called the generalized Hurst exponent. The Hurst exponent is measure of self-similarity and correlation properties of time series produced by fractal. The presence or absence of long range correlation can be determined using Hurst exponent. A monofractal time series is characterized by unique h(q) for all values of q.

The generalized Hurst exponent h(q) of MFDFA is related to the classical scaling exponent $\tau(q)$ by the relation

$$\tau(q) = qh(q) - 1 \quad (4)$$

A monofractal series with long range correlation is characterized by linearly dependent q order exponent $\tau(q)$ with a single Hurst exponent H. Multifractal signal on the other hand, possess multiple Hurst exponent and in this case, $\tau(q)$ depends non-linearly on q [9].

The singularity spectrum f(α) is related to h(q) by

$$\alpha = h(q) + qh'(q)$$
$$f(\alpha) = q[\alpha - h(q)] + 1$$

Where α denoting the singularity strength and *f(α)*, the dimension of subset series that is characterized by α. The width of the multifractal spectrum essentially denotes the range of exponents. The spectra can be characterized quantitatively by fitting a quadratic function with the help of least square method [9] in the neighbourhood of maximum $\alpha_0$,

$$f(\alpha) = A(\alpha - \alpha_0)^2 + B(\alpha - \alpha_0) + C \quad (5)$$

Here C is an additive constant C = f($\alpha_0$) = 1 and B is a measure of asymmetry of the spectrum. So obviously it is zero for a perfectly symmetric spectrum. We can obtain the width of the spectrum very easily by extrapolating the fitted quadratic curve to zero.
Width W is defined as,

$$W = \alpha_1 - \alpha_2 \quad (6)$$

with $f(\alpha_1) = f(\alpha_2) = 0$

The width of the spectrum gives a measure of the multifractality of the spectrum. Greater is the value of the width W greater will be the multifractality of the spectrum. For a monofractal time series, the width will be zero as h(q) is independent of q. The spectral width has been considered as a parameter to evaluate how a group of string instruments vary in their pattern of playing from another

**RESULTS AND DISCUSSION:**

Musical structures can be explored on the basis of multifractal analysis and nonlinear correlations in the data. Traditional signal processing techniques are not capable of identifying such relationships, nor do they provide quantitative measurement of the complexity or information content in the signal. The following figures (**Fig. 2a-f**) show quantitatively how the complexity patterns of each *raga* clip vary significantly from the others giving a cue for different levels of emotional arousal corresponding to each clip.

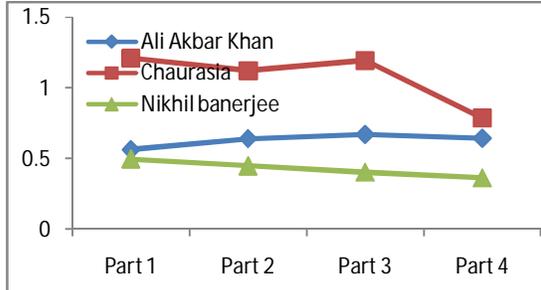
Fig. 2a: Variation of multifractal width within *raga Hamswadhani*

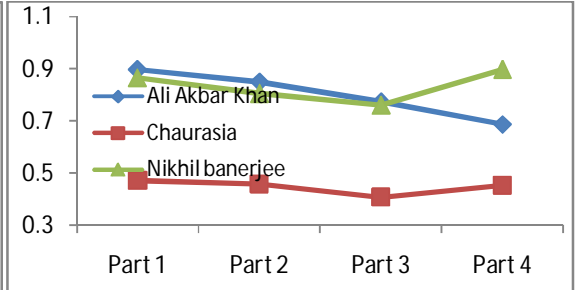
Fig. 2b: Variation of multifractal width within *raga Darbari*

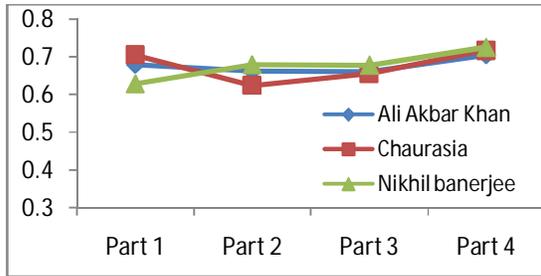
Fig. 2c: Variation of multifractal width within *raga Jai jwanti*

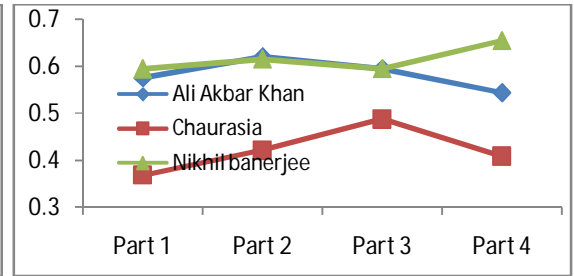
Fig. 2d: Variation of multifractal width within *raga Mian ki..*

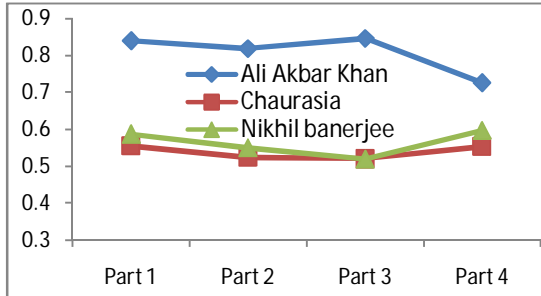
Fig. 2e: Variation of multifractal width within *raga Durga*

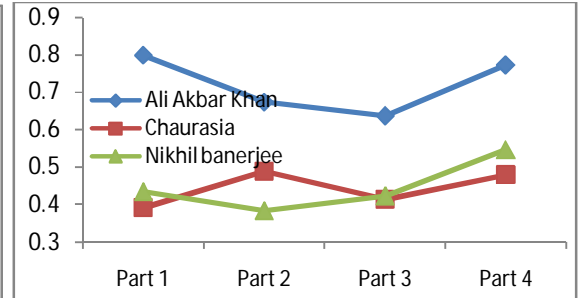
Fig. 2f: Variation of multifractal width within *raga Yaman*

We see that in most cases the variation of multifractal widths within a particular *raga* is almost similar for all the artistes; though the characteristic values of multifractal widths are distinctly different from one clip to other. The similarity in fluctuation patterns within each *raga* may be attributed to the strict intonation pattern followed by all the artistes during the performance of a *raga;* while the difference in the characteristic values may be a signature of artistic style. Also, in many parts we find that an artist has deviated significantly from the characteristic pattern of that *raga;* herein lies the cue for artistic improvisation where the artist uses his own creativity to create something new from the obvious structure of *raga*. In **fig. 2b**, we see that in the last part the complexity value significantly increasing for the *sitar* clip as opposed to the *sarod* clip; while in **Fig. 2f** we find that in the 2nd part complexity value dipping for the *flute* clip as opposed to the other two clips where the complexity values are increasing. In this way, we can have an estimate of how much an artist improvises during the rendition of a particular *raga*.

The averaged values for each raga clips have been given in the following table (**Table 1**) and the corresponding fig (**Fig. 3**) shows the values for each artist:

|  | **Hamswadhani** | **Darbari** | **Jay Jayanti** | **Mia ki Malhar** | **Durga** | **Yaman** |
|---|---|---|---|---|---|---|
| **Ali Akbar Khan** | 0.628 | 0.802 | 0.677 | 0.583 | 0.809 | 0.721 |
| **Chaurasia** | 1.076 | 0.447 | 0.675 | 0.421 | 0.538 | 0.443 |
| **Nikhil banerjee** | 0.429 | 0.647 | 0.403 | 0.614 | 0.564 | 0.447 |

Table 1: Variation of multifractal width corresponding to *ragas* of contrast emotion by different artistes

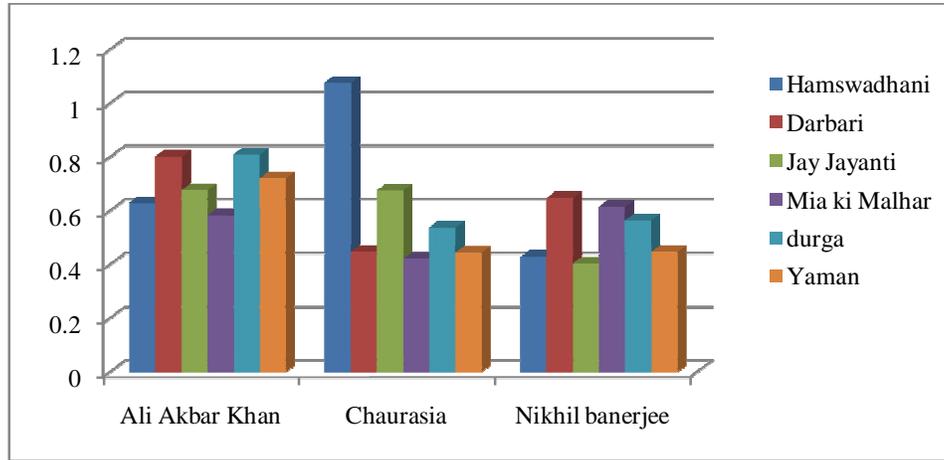

**Fig. 3: Clustering of multifractal widths for each artist corresponding to each *raga***

From the above figure it is clear that there is distinct categorization of emotional responses corresponding to each *raga* clip. In case of *sarod* and *sitar*, we find that *raga Hamswadhani* (corresponding to happy emotion) has a lower value of complexity as opposed to the *flute* clip where the complexity value is significantly high. The complexity values corresponding to *raga Darbari* (depicting sad emotion) is consistently high for *sarod* and *sitar* while that is significantly low for *flute* clip. In case of the other pair *Jaijwanti* (happy clip) and *Mia ki Malhar* (sorrow clip), we see that there is similarity in response for *sarod* and *flute,* i.e. complexity values on the higher side for happy clip while it is lower for sad clip; the response is vice-versa for *sitar* clip. In case of the other pair, i.e. raga *Durga* (mainly on the happier side but is mixed with other emotions like romance, serene etc.) and *raga Yaman* (mainly on the negative side of Russel's emotional sphere but is mixed with other emotions like devotion etc.) there was considerable ambiguity even when it comes to human response psychological data. The same has been reflected in our results where the average difference in complexity of these two *ragas* is not so significant as compared to the other two pairs. Our study thus points in the direction of timbre specific categorization of emotion in respect to *Hindustani raga* music. We see that the emotion classification works the best for *flute* where the difference in complexity for the happy and sad clips is the maximum; while the difference is minimum for *sarod*, thus it is difficult to categorize emotions from *sarod* clips.

Thus, in this work we have developed an automated emotion classification algorithm with which we can quantify and categorize emotions corresponding to a particular instrument. Also, the complexity values give a hint for style recognition corresponding to a particular artist.

## CONCLUSION:

This study presents a first-of-its kind data in regard to categorization and quantification of emotional arousal based responses to Hindustani classical music. The inherent ambiguities said to be present in Hindustani classical music is also reflected beautifully in the results. That a particular *raga* can portray an amalgamation of a number of perceived emotions can now be tagged with the rise or fall of multifractal width or complexity values associated with that *raga*. The study presents the following interesting conclusions which have been listed below:

1. For the first time, an association have been made with the timbre of a particular instrument with the variety of emotions that it conveys. Thus for effective emotional classification, timbre of the instrument will play a very important role in future studies.
2. The multifractal spectral width has been used as a timbral parameter to quantify and categorize emotional arousal corresponding to a particular clip played in a specific instrument.
3. We try to develop a threshold value for a particular instrument using multifractal spectral width, beyond which emotions will change. The following figures (**Fig. 4**) summarizes the results:

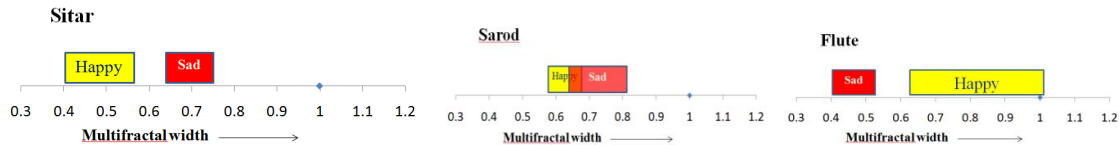

Fig. 3: Use of multifractal width as a tool to categorize emotions in different instruments

(i) From the plot it is clear that emotional classification can be best done with the help of flute where the complexity values of happy and sad clips are distinctly different from one another.

(ii) There is an overlap in case of *sarod* clips between happy and sad complexity values. This can be attributed to the inherent ambiguity present in the clips of *Hindustani classical music,* i.e there cannot be anything as complete joy or complete sorrow, there remains always states which are between joy and sorrow, which is beautifully reflected in the overlap part of the two emotions.

In conclusion, this study provides a novel tool and a robust algorithm with which future studies in the direction of emotion categorization using music clips can be carried out keeping in mind the timbral properties of the sound being used. A detailed study using a variety of other instruments and *ragas* is being carried out to yield more conclusive result.

## ACKNOWLEDGEMENT:

One of the authors, AB acknowledges the Department of Science and Technology (DST), Govt. of India for providing (A.20020/11/97-IFD) the DST Inspire Fellowship to pursue this research work. The first author, SS acknowledges the West Bengal State Council of Science and Technology (WBSCST), Govt. of West Bengal for providing the S.N. Bose Research Fellowship Award to pursue this research (193/WBSCST/F/0520/14).